\documentclass[
  aps,
  pra,
  preprint,
  onecolumn,
  groupedaddress,
  nofootinbib
]{revtex4-2}

\usepackage[T1]{fontenc}
\usepackage{microtype}
\usepackage[hidelinks]{hyperref}
\usepackage{graphicx} % Required for inserting images
\usepackage{array} \usepackage{hyperref,amsmath,amsthm,amssymb, graphicx, array, bbm, enumitem, physics} \usepackage{cleveref} \usepackage{lineno} \usepackage{titlesec} \usepackage{comment} \usepackage{minted}          \newcommand{\bq}{\mathbf{q}}        \newcommand{\hp}{\hat{\phi}}  \newcommand{\hs}{\hat{\sigma}}   \newcommand{\R}{\mathbb{R}}       
 
  \usepackage{float} \usepackage{longtable}         \usepackage{physics} \usepackage{quantikz}
\usepackage{cancel}

\newcommand{\xctrl}[1]{%
  \control[style={
    minimum size=4pt,
    inner sep=0pt,
    append after command={
      (\tikzlastnode.north)
      node[above,inner sep=1pt,font=\scriptsize]{X}
    }
  }]{}%
  \ifdefined\vqw
    \vqw{#1}%
  \else
    \wire[d][#1]{q}%
  \fi
}

\newcommand{\xmeter}{%
  \meter{}%
  \makebox[0pt][l]{%
    \hspace{.8em}%
    \raisebox{-0.9em}[0pt][0pt]{$\scriptsize X$}%
  }%
}

\makeatletter
\def\@hangfrom@section#1#2#3{\@hangfrom{#1#2}{#3}}
\def\@hangfroms@section#1#2{#1#2}
\makeatother

\begin{document}

\title{Chiral waveguide model for error-correctable quantum computation}

\author{Zixiang Hong}
\thanks{\href{mailto:tom.hong@yale.edu}{tom.hong@yale.edu}}
\affiliation{%
Department of Applied and Computational Mathematics,\\
Yale University, New Haven, Connecticut 06511, USA}

% Suppress both the date and REVTeX's “Dated:” label.
\date[]{}

\begin{abstract}
We develop a first-principles framework connecting microscopic waveguide scattering to quantum error correction and algorithmic performance. For polarization-encoded single photons interacting with an ensemble of two-level atoms in a chiral waveguide, we derive an effective polarization quantum channel by tracing out the frequency degree of freedom. Using this channel as the physical noise model for a rotated planar code, we observe threshold behavior as the number of atoms $N$ and the atom-waveguide coupling strength $\Gamma$ are varied. An asymptotic analysis of the microscopic scattering model explains the threshold scaling with $N\Gamma/\sigma$, where $\sigma$ is the photon spectral bandwidth. For identical, lossless atoms with symmetric real coupling and a photon spectrum centered at resonance, the effective noise is strongly biased toward $X$ errors. We then incorporate the resulting logical errors into quantum phase estimation and demonstrate that the noise bias can guide circuit design. In particular, a Hadamard-basis implementation reduces sensitivity to the dominant error, improving algorithmic performance.
\end{abstract}

\maketitle

\clearpage

\section{Introduction}
An important challenge in quantum information processing is connecting the microscopic physics of a hardware platform to the performance of an error-corrected quantum algorithm \cite{Iyer_2018,Chamberland2022}. Microscopic models describe the interactions that generate noise \cite{Nysteen2017,Puri2020,Ozeri2007,Schrinski2022}; quantum channels characterize its effect on quantum information \cite{Rahn2002,Kesting2013}; quantum error correction determine the resulting logical error rates \cite{Fowler2012,Beale2018,Stephens2014,BonillaAtaides2021}; and algorithm-level simulations illustrate the impact of logical errors on algorithmic performance \cite{Dutkiewicz2025,Liang2024,Suzuki2022}. Some of these levels have been connected by mapping physical noise channels to effective logical channels \cite{Darmawan2021} and analyzing algorithmic performance under logical noise \cite{Chamberland2022,Ding2023,Ding2026}. In particular, error models used in error correction and algorithm-level studies are often specified phenomenologically, for example using Lindblad master equations, rather than derived directly from the microscopic Hamiltonian of the underlying platform \cite{Darmawan2021,Rahn2002,Chamberland2022}. Conversely, carrying a detailed microscopic model through error correction to an algorithmic output requires additional assumptions about the error correction procedure and the location of logical noise. Here, we examine this connection for a specific waveguide scattering model, error correction code, and quantum algorithm.

Establishing this connection offers two benefits. First, within the assumptions of the modeling chain, it relates experimentally relevant parameters to logical and algorithmic performance. Rather than specifying an abstract error probability, one can examine how physical parameters such as coupling strength, spectral bandwidth, disorder, and the number of scattering elements affect logical error rates and the accuracy of a quantum computation \cite{Darmawan2021,Levy-Yeyati2025}. This connection can also inform resource estimates by relating physical parameters to the error correction overhead needed to achieve a target algorithmic accuracy. Second, a microscopic derivation can determine features of the noise that may be absent from simplified phenomenological models unless they are included explicitly. These features, including error bias and dependence on physical parameters, can motivate comparisons between error correction codes and modifications to the implementation of a quantum algorithm \cite{Fellous-Asiani2025}.

In this work, we develop a proof-of-concept framework linking a chiral waveguide model to error correction and algorithmic performance. Starting from the microscopic Hamiltonian, we analyze single-photon scattering by an ensemble of two-level atoms and derive an effective quantum channel for the polarization qubit by accounting for the finite bandwidth $\sigma$. We then use this channel as the physical noise model for a rotated planar code. For identical, lossless atoms with symmetric, real coupling, we find a threshold boundary governed by $N\Gamma/\sigma$ where $N$ is the number of atoms and $\Gamma$ is the atom-waveguide coupling strength. In the regime $N\Gamma \ll \sigma$, an asymptotic analysis of the scattering model reproduces this scaling and shows that the effective Pauli noise is strongly biased toward $X$ errors in the resonant, lossless limit. We then propagate the resulting logical errors into quantum phase estimation, relating its performance directly to the coupling strength and number of atoms. This $X$ bias also motivates a Hadamard-basis implementation of phase estimation that reduces sensitivity to the dominant error at selected transmission locations. 

The present analysis is not yet a complete end-to-end model of a quantum device. A device-level description would also require microscopic models of state preparation, single-qubit and two-qubit gates, ancilla preparation and readout, and final logical measurement, including their imperfections and possible correlations. Many of these components have been studied separately in waveguide-based systems \cite{Appel2021,Paulisch_2016,Mahmoodian2016,Malz2020,Cheng2023}. The framework developed here thus provides a starting point for exploring their integration. Further extensions could incorporate photon loss, nonidentical atoms, random variations in coupling strengths and atomic positions, and other experimentally relevant imperfections. More broadly speaking, an analogous modeling strategy could be applied to other platforms, including circuit QED \cite{OBrien2017}, neutral-atom \cite{Wu2022}, and free-space photonic systems \cite{Gorshkov2011}. In these platforms, microscopic models can similarly be used to derive effective noise channels and assess their effects at the logical and algorithmic levels.

This paper is organized as follows. In \cref{sec:WaveguideModel}, we introduce the chiral waveguide model and derive the single-photon scattering matrix from the microscopic Hamiltonian. In \cref{sec:Communicationchannel}, we account for the photon's finite spectral bandwidth to obtain an effective quantum channel for its polarization qubit. In \cref{sec:Rotatedplanarcode}, we use this channel as the physical noise model to study error correction with the rotated planar code. In \cref{sec:QuickIntroductionanderrormodel1}, we review the rotated planar code using the stabilizer formalism and describe how physical errors are introduced into the code. In \cref{sec:Thresholds}, we identify threshold behavior as the coupling strength or number of atoms varies. In \cref{sec:BoundaryMap}, we present a boundary map of these thresholds and use asymptotic analysis to relate the numerical boundary to the microscopic model. In \cref{sec:QuantumPhaseEstimateAlgorithm}, we incorporate the resulting logical errors into quantum phase estimation. In \cref{sec:Quickintroductionanderrormodel2}, we review the phase estimation protocol and describe how logical errors are modeled in the circuit. In \cref{sec:results}, we examine how phase estimation performance depends on the waveguide parameters and the number of ancilla qubits. In \cref{sec:OneImprovement}, we use the error bias derived from the microscopic model to develop a Hadamard-basis implementation of phase estimation and compare it with the standard implementation. Finally, in \cref{sec:discussion}, we summarize the results and discuss extensions toward a more complete end-to-end model.

\section{Waveguide model}
\label{sec:WaveguideModel}
We consider a single chiral waveguide supporting two polarization modes that interact with an ensemble of $N$ identical two-level atoms. The total Hamiltonian consists of three terms, $\hat H=\hat H_F+\hat H_A+\hat H_I$ where $\hat H_F$, $\hat H_A$, and $\hat H_I$ describe the photon field, atoms, and their interaction, respectively. Under real-space quantization, the photon field Hamiltonian is
\begin{equation}
\hat H_F
=\hbar\sum_{\alpha=\pm}\int dx\,
\hat\phi_\alpha^\dagger(x)
\left[\omega_0-i v_g\frac{\partial}{\partial x}\right]
\hat\phi_\alpha(x),\\
\end{equation}
where $\omega_0$ is the reference frequency about which the waveguide dispersion is linearized, and $v_g$ is the photon group velocity. The operator $\hp_\alpha^\dagger(x)$ creates a photon with polarization $\alpha$ at position $x$ and satisfies the canonical commutation relation
\begin{equation}
\left[\hat\phi_\alpha(x),\hat\phi_\beta^\dagger(x')\right]
=\delta_{\alpha\beta}\delta(x-x').
\end{equation}
The atomic Hamiltonian is given by
\begin{equation}
\hat H_A
=\hbar\sum_{j=1}^{N}\Omega
\hat\sigma_j^\dagger\hat\sigma_j,\\
\end{equation}
where $\Omega$ is the transition frequency of the $N$ identical, lossless atoms. The operator $\sigma_j^\dagger$ creates an atomic excitation for $j$-th atom, located at $x_j$, and obeys the anticommutation relation onsite 
\begin{equation}
\{\hs_j,\hs_j^\dagger \} = 1, \qquad  \{\hs_j,\hs_j \} = 0,
\end{equation}
and obeys the commutation relations offsite
\begin{equation}
[\hs_i,\hs_j^\dagger] = 0, \qquad [\hs_i,\hs_j]=0, \qquad    \text{when $i \not= j$.}
\end{equation}
Under rotating-wave approximation, the interaction Hamiltonian is described by 
\begin{equation}
\hat H_I
=\hbar\sum_{j=1}^{N}\sum_{\alpha=\pm}\int dx\,
\delta(x-x_j)\left[V_{\alpha}\hat\phi_\alpha^\dagger(x)\hat\sigma_j
+\mathrm{H.c.}\right],
\end{equation}
where $V_{\alpha}$ is the evanescent coupling strength between atoms and the waveguide continuum with polarization $\alpha$.

We consider a general state in the single-excitation sector,
\begin{equation}
\ket{\Psi}
=\sum_{\alpha=\pm}\int dx\,\psi_\alpha(x)\hat\phi_\alpha^\dagger(x)\ket{\varnothing}
+\sum_j a_j\hat\sigma_j^\dagger\ket{\varnothing},
\end{equation}
where $\ket{\varnothing}$ is the combined state of vacuum state of the photon field and ground state of the atoms. The function $\psi_\alpha(x)$ denotes the probability amplitude of the photon presenting at location $x$ in $\alpha$ polarization and $a_j$ is the probability amplitude for exciting the $j$-th atom. 
The state satisfies the time-independent Schr\"odinger equation $\hat H\ket{\Psi}=\hbar\omega\ket{\Psi}$, where $\omega$ is the frequency of the photon. By projecting both sides to the states $\phi_\alpha^\dagger(x)\ket{\varnothing}$ and $\sigma_j^\dagger\ket{\varnothing}$, we have the following system of equations, for $\alpha = \pm$,
\begin{equation}
-i v_g\frac{\partial\psi_\alpha(x)}{\partial x}
+\sum_{j=1}^{N}V_{\alpha}a_j\delta(x-x_j)
=(\omega-\omega_0)\psi_\alpha(x),
\end{equation}
\begin{equation}
\sum_{\beta=\pm}V_{\beta}^{*}\psi_\beta(x_j)
=(\omega-\Omega)a_j.
\end{equation}
By eliminating the atomic amplitude $a_j$ from the above system, we have
\begin{equation}
\label{eq:MainEquation}
-i v_g\frac{\partial\psi_\alpha(x)}{\partial x}
+\sum_{j=1}^{N}\sum_{\beta=\pm}
\frac{V_{\alpha}V_{\beta}^{*}}{\omega-\Omega}
\psi_\beta(x_j)\delta(x-x_j)
=(\omega-\omega_0)\psi_\alpha(x),
\end{equation}
where $\alpha = \pm$. Away from the atomic positions $x\neq x_j$, the solutions of \cref{eq:MainEquation} are plane waves of the form $\psi_\alpha(x)=Ae^{iqx}$ where $A$ is the amplitude needed to be fixed and $q=\frac{\omega-\omega_0}{v_g}$. Consider ansatz 
\begin{equation}
\psi_\alpha(x)=
\begin{cases}
t_{0\alpha}e^{iqx}, & x<x_1,\\
t_{1\alpha}e^{iqx}, & x_1\le x\le x_2,\\
\vdots & \\
t_{N\alpha}e^{iqx}, & x>x_N,
\end{cases}
\end{equation}
where the input amplitudes are normalized as $|t_{0+}|^2+|t_{0-}|^2=1$. To determine the amplitudes $t_{j\alpha},\ldots,t_{N\alpha}$, we integrate \cref{eq:MainEquation} over an infinitesimal interval
$[x_j-\epsilon,x_j+\epsilon]$ around each atom: 
\begin{equation}
-i v_g\left[\psi_\alpha(x_j+\epsilon)-\psi_\alpha(x_j-\epsilon)\right]
+\sum_{\beta=\pm}\frac{V_{\alpha}V_{\beta}^{*}}
{\omega-\Omega}\psi_\beta(x_j)=0.
\end{equation}
We regularize the discontinuity in $\psi_\alpha$ using
\begin{equation}
\psi_\alpha(x_j)=\lim_{\epsilon\to0}
\frac{\psi_\alpha(x_j+\epsilon)+\psi_\alpha(x_j-\epsilon)}{2}.
\end{equation}
Then, we have
\begin{equation}
i v_g\left[t_{j\alpha}-t_{(j-1)\alpha}\right]
=\frac{1}{2}\sum_{\beta=\pm}
\left[t_{j\beta}+t_{(j-1)\beta}\right].
\end{equation}
Solving these equations yields the single-atom transfer matrix $T$, which relates the amplitudes on either side of the $j$th atom:
\begin{equation}
\begin{pmatrix}
t_{j+}\\ t_{j-}
\end{pmatrix}
=T
\begin{pmatrix}
t_{(j-1)+}\\ t_{(j-1)-}
\end{pmatrix},
\end{equation}
where 
\begin{equation}
\label{eq:IndividualTransferMatrix}
T = \frac{1}{\Delta+i(\Gamma_{+}+\Gamma_{-})}
\begin{pmatrix}
\Delta+i(\Gamma_{-}-\Gamma_{+})
&-i\dfrac{V_{+}V_{-}^{*}}{v_g}\\
-i\dfrac{V_{-}V_{+}^{*}}{v_g}
&\Delta+i(\Gamma_{+}-\Gamma_{-})
\end{pmatrix},
\end{equation}
with 
\begin{equation}
\Delta=\omega-\Omega,
\end{equation}
\begin{equation}
\Gamma_{\alpha}=\frac{|V_{\alpha}|^2}{2v_g}.
\end{equation}
The total transfer matrix $U$, relating the input and output amplitudes, is defined by
\begin{equation}
\begin{pmatrix}
t_{N+}\\ t_{N-}
\end{pmatrix}
=U
\begin{pmatrix}
t_{0+}\\ t_{0-}
\end{pmatrix},
\end{equation}
where 
\begin{equation}
U = T^N.
\end{equation}
For lossless atoms, $T$ is unitary, ensuring conservation of the total polarization probability, $\abs{t_{N+}}^2 + \abs{t_{N-}}^2 = 1$. Furthermore, up to a global phase, $T$ can be expressed as a single-qubit rotation,
\begin{equation}
T = e^{-i\theta} R_{\hat{n}}(2\theta),
\end{equation}
where the rotation angle is
\begin{equation}
\theta=\arctan\left(\frac{\Gamma_{+}+\Gamma_{-}}{\omega-\Omega}\right) 
\end{equation}
rotation axis is
\begin{equation}
\hat{n} = \frac{1}{|V_{+}|^2+|V_{-}|^2}
\begin{bmatrix}
2\operatorname{Re}\bigl(V_{+}^{*}V_{-}\bigr)\\
2\operatorname{Im}\bigl(V_{+}^{*}V_{-}\bigr)\\
|V_{+}|^2-|V_{-}|^2
\end{bmatrix},
\end{equation}
and 
\begin{equation}
R_{\hat{n}}(\theta) = \exp[-\frac{i\theta}{2}\hat{n} \cdot\hat{\sigma}],
\end{equation}
where $\hat{\sigma} = \{\sigma_x,\sigma_y,\sigma_z\}$ and detailed derivation is in Appendix \ref{appendix:Arbitraryqubitrotation}.

The same model can also describe two chiral waveguide modes using dual-rail encoding.

\section{Effective polarization quantum channel}
\label{sec:Communicationchannel}
We treat polarization and frequency as separate degrees of freedom and define the single-photon state
\begin{equation}
\ket{+,\omega}=\ket{+}\otimes\ket{\omega} = \frac{1}{\sqrt{v_g}}a_{+}^{\dagger}(\omega)\ket{0}
\end{equation}
with the normalization
\begin{equation}
\langle\alpha,\omega|\alpha',\omega'\rangle=\delta_{\alpha\alpha'}\delta(\omega-\omega'),
\end{equation}
which follows from the dispersion relation
\begin{equation}
\omega=v_g k+\omega_0,\qquad k=\frac{\omega-\omega_0}{v_g}.
\end{equation}
We consider a Gaussian spectral envelope centered at $\omega_{center}$
\begin{equation}
f(\omega)=\frac{1}{\sigma^{1/2}\pi^{1/4}}\exp\left[-\frac{(\omega-\omega_{center})^2}{2\sigma^2}\right],
\end{equation}
where $\sigma$ characterizes the bandwidth and $f(\omega)$ is normalized according to
\begin{equation}
\int_{-\infty}^{\infty}d\omega\,|f(\omega)|^2=1.
\end{equation}
The corresponding frequency wavepacket is
\begin{equation}
\ket{f}=\int d\omega\,f(\omega)\ket{\omega}.
\end{equation}
We assume that polarization and frequency are initially uncorrelated, so the input state can be written as
\begin{equation}
\rho_{\mathrm{in}}=\rho_{\mathrm{pol}}\otimes\ket{f}\bra{f}.
\end{equation}
In the lossless single-photon scattering model considered here, the photon frequency is conserved. The full scattering unitary can therefore be written as
\begin{equation}
S_N=\int d\omega\,U_N(\omega)\otimes\ket{\omega}\bra{\omega},
\end{equation}
where
\begin{equation}
U_N(\omega)=T^N(\omega),
\end{equation}
is the total transfer matrix at frequency $\omega$ for $N$ atoms.
The joint output state is then
\begin{equation}
\begin{split}
\rho_{\mathrm{joint,out}}&=S_N\bigl(\rho_{\mathrm{pol}}\otimes\ket{f}\bra{f}\bigr)S_N^{\dagger}\\
&= \int d\omega\int d\omega'\,U_N(\omega)\rho_{\mathrm{pol}}U_N^{\dagger}(\omega')
\otimes\ket{\omega}\bra{\omega'}f(\omega)f^{*}(\omega').
\end{split}
\end{equation}
Tracing out the frequency degree of freedom yields the reduced polarization state
\begin{equation}
\begin{split}
\rho_{\mathrm{out}}&=\operatorname{Tr}_{\omega}\left[\rho_{\mathrm{joint,out}}\right]\\
&= \int d\omega\,U_N(\omega)\rho_{\mathrm{pol}}U_N^{\dagger}(\omega)|f(\omega)|^2.
\end{split}
\end{equation}
We thus define the effective quantum channel acting on the polarization qubit as
\begin{equation}
\label{eq:MainQuantumChannel}
\mathcal N[\rho]=\int d\omega\,U_N(\omega)\rho U_N^{\dagger}(\omega)|f(\omega)|^2.
\end{equation}

\section{Rotated planar code}
\label{sec:Rotatedplanarcode}
We consider the task of transmitting a qubit between spatially separated locations through the waveguide system. Frequency-dependent atom-photon scattering can introduce errors in the polarization state during transmission. To suppress these errors, we encode the qubit using a rotated planar code and investigate its error correction performance under the effective polarization channel derived in the preceding section.

The rotated planar code is a variant of the surface code obtained by rotating the conventional planar-code lattice by $45$ degree, reducing the number of physical qubits required for a given code distance. A distance-$d$ rotated planar code encodes one logical qubit in $d^2$ physical qubits and can correct up to $\lfloor (d-1)/2 \rfloor$ arbitrary physical-qubit errors. We model each physical qubit as the polarization state of a single photon propagating through an individual chiral waveguide. During transmission, each physical qubit is subject to the scattering-induced noise described by the effective polarization channel.
\subsection{Error model}
\label{sec:QuickIntroductionanderrormodel1}
We review the rotated planar code within the stabilizer formalism, where the code space is defined as the simultaneous $+1$ eigenspace of a set of mutually commuting Pauli operators. A distance-$d$ rotated planar code encodes one logical qubit into $d^2$ physical qubits arranged on a square lattice with open boundaries. The stabilizer generators consist of bulk and boundary stabilizers. Each bulk stabilizer acts with either four $X$ or four $Z$ operators on the vertices of a plaquette, with $X$-type and $Z$-type stabilizers arranged in a checkerboard pattern. Each boundary stabilizer acts on two neighboring boundary qubits. The $X$-type boundary stabilizers lie on two opposite edges, while the $Z$-type stabilizers lie on the other two. The stabilizers are arranged so that the supports of any $X$-type and $Z$-type stabilizers overlap on either zero or two qubits, ensuring that they commute. Further details can be found in \cite{Horsman2012}.

For simulations at large code distances, we replace the original quantum channel with its Pauli-twirled counterpart, restricting the noise to be Pauli errors. The Pauli-twirled channel corresponding to \cref{eq:MainQuantumChannel} is
\begin{equation}
\mathcal N_{PT}[\rho]=\sum_{P\in \{I, X, Y, Z\}}p_P P\rho P,
\end{equation}
where the probability of a Pauli error $P$ is
\begin{equation}
\label{eq:PauliErrorProbability}
p_P=\int d\omega\,|f(\omega)|^2|u_P(\omega)|^2
\end{equation}
with $P\in\{I,X,Y,Z\}$, and 
\begin{equation}
u_P(\omega)=\frac{1}{2}\operatorname{Tr}\left(PU_N(\omega)\right).
\end{equation}
The derivation of this channel and the conditions under which Pauli twirling accurately approximates the original channel are discussed in Appendix \ref{appendix:Pauli-twirlingchannel}.  For symmetric real coupling and a Gaussian photon spectrum centered at the atomic resonance, the coherent contributions vanish under spectral averaging, so the original channel coincides exactly with its Pauli-twirled counterpart. Since the resulting noise is a Pauli channel, Clifford-based error correction simulations reproduce the same logical error statistics as exact quantum-state simulations under the same idealized error correction operations.

In our simulations, we assume ideal state preparation, ancilla control, and syndrome measurements, so the only modeled errors arise from photon transmission through the waveguide system. Under these assumptions, we perform a single round of syndrome extraction and decode the measured syndrome using minimum-weight perfect matching \cite{Dennis2002,Fowler2012}.

Although we assume ideal error correction operations, several of the required components have been investigated in waveguide-based systems. Our waveguide model describes single-qubit polarization rotations, while controlled-$Z$ gates have been demonstrated using three-level atomic systems \cite{Schrinski2022} and two-level emitters in chiral multimode waveguides \cite{LevyYeyati2025}. Polarization readout could be implemented using a polarizing beam splitter and waveguide-integrated superconducting nanowire single-photon detector (SNSPD) arrays with spatial and temporal multiplexing. Time-resolved detection can provide photon-number and detection-position information \cite{Cheng2023}. These developments provide a basis for incorporating realistic gate and measurement imperfections into future extensions of our framework. We present several case studies below.

\subsection{Thresholds for fixed coupling strength and fixed number of atoms}
\label{sec:Thresholds}
We consider identical, lossless atoms with symmetric real coupling strengths, $V_+=V_-=V\in \R$, and assume that the atomic transition frequency is resonant with the central photon frequency, $\Omega=\omega_{center}$.
\begin{figure}[ht!]
\centering

    \noindent\makebox[0.5\linewidth][l]{\large\bfseries (a)}
    \vspace{0.4em}

    \includegraphics[width=0.5\linewidth]{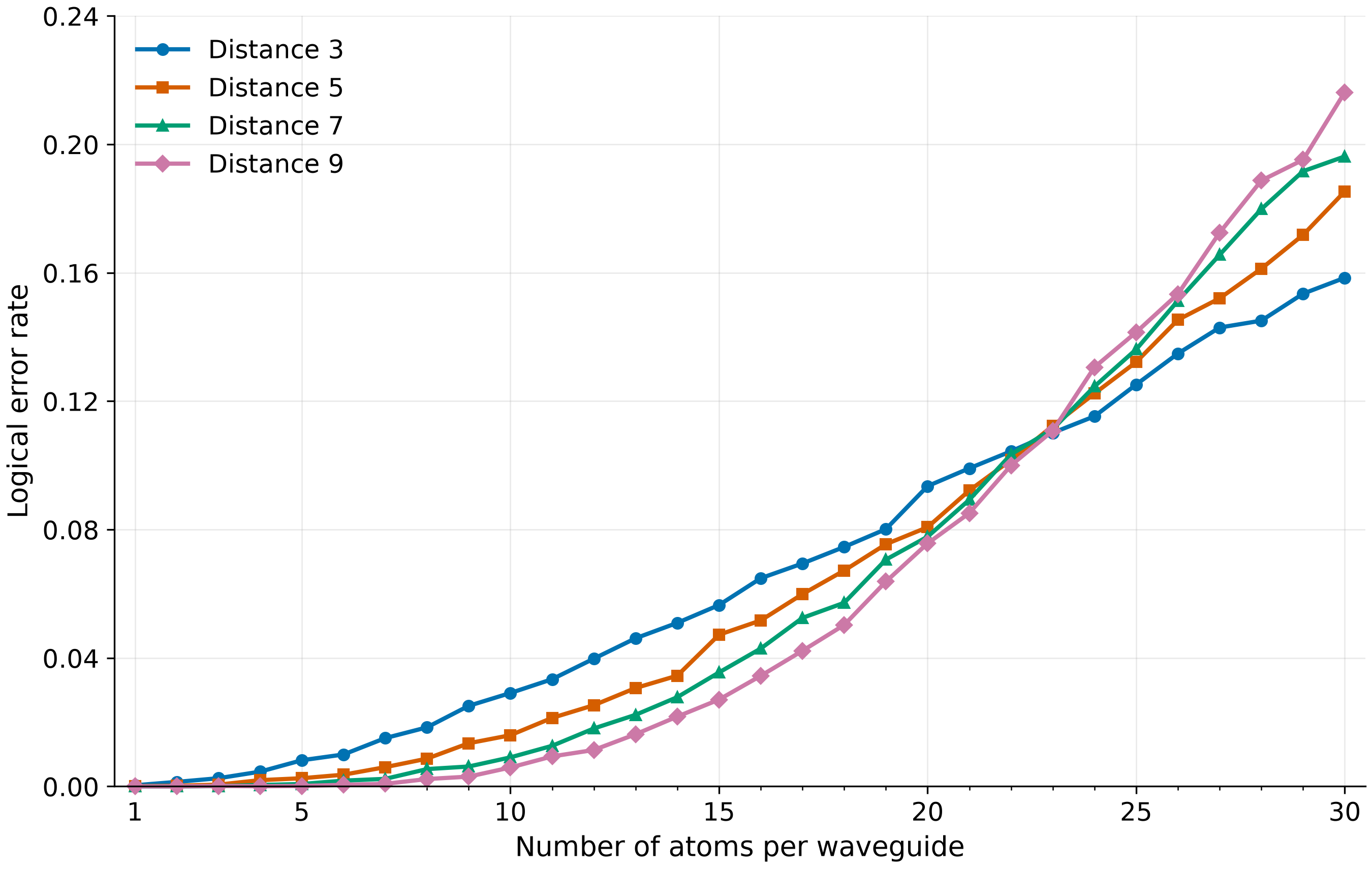}

    \vspace{0.4em}

    \noindent\makebox[0.5\linewidth][l]{\large\bfseries (b)}
    \vspace{0.4em}

    \includegraphics[width=0.5\linewidth]{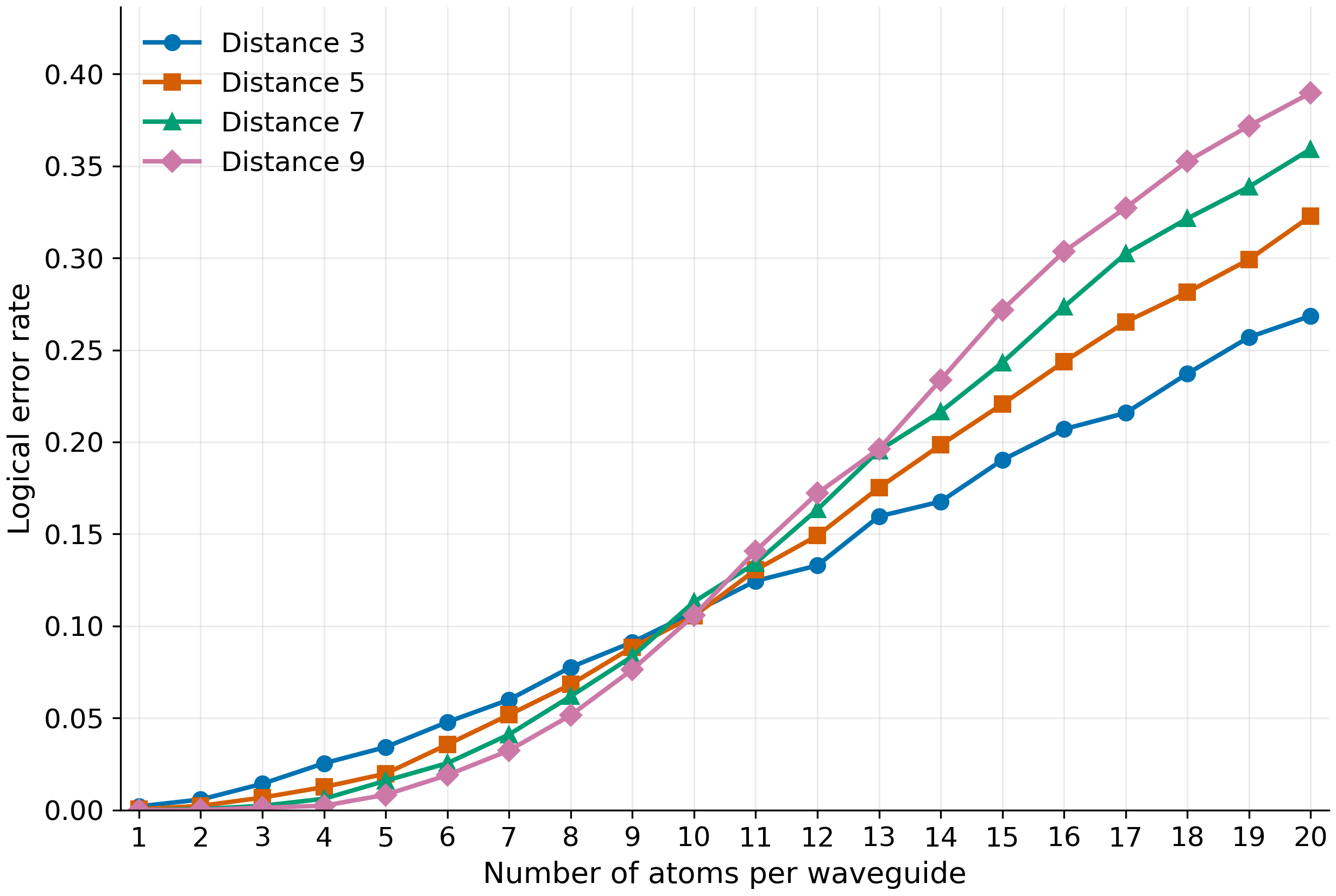}

    \vspace{0.4em}

    \noindent\makebox[0.5\linewidth][l]{\large\bfseries (c)}
    \vspace{0.4em}

    \includegraphics[width=0.5\linewidth]{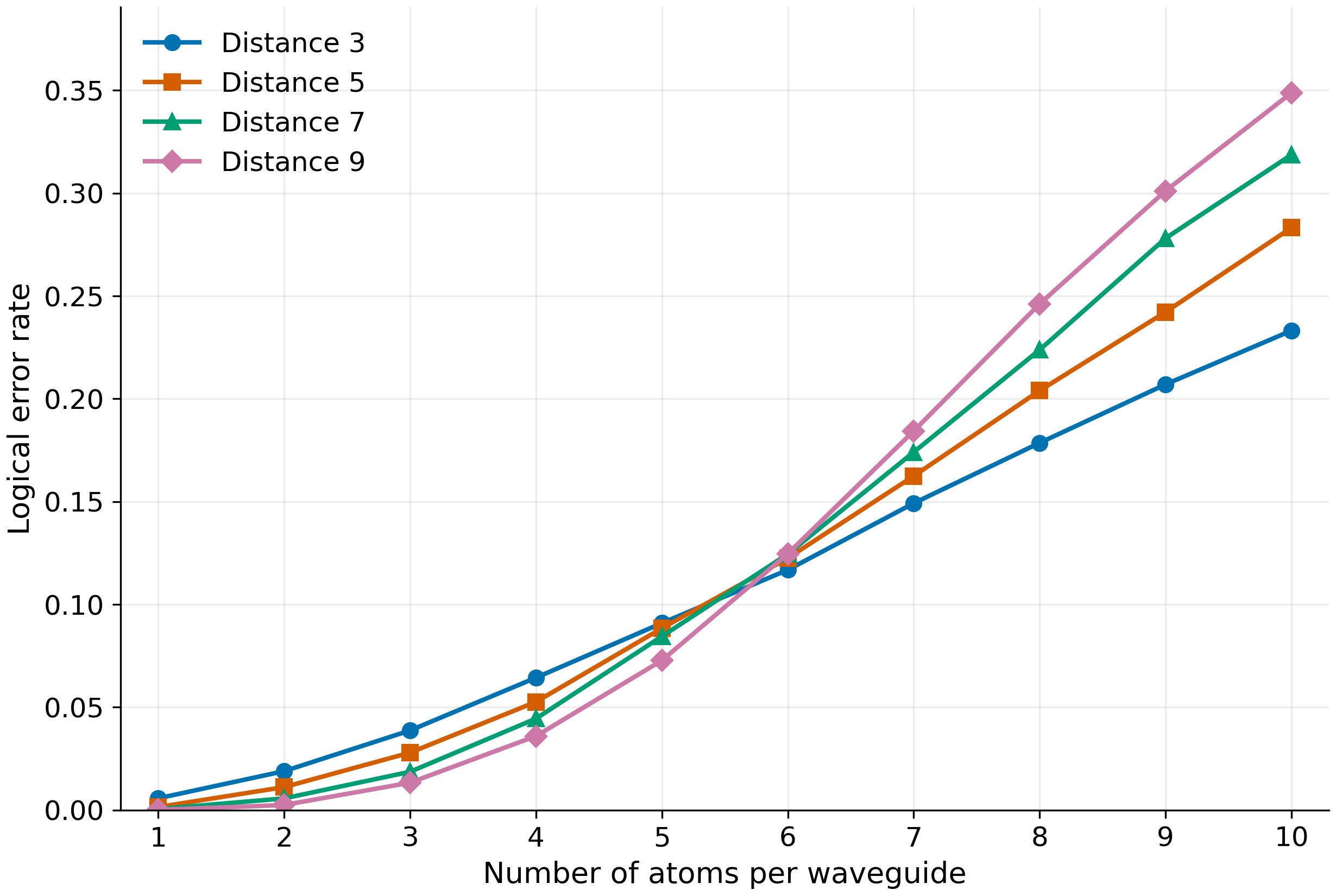}

    \caption{Logical error rate as a function of the number of atoms $N$ for fixed coupling strengths (a) $\Gamma/\sigma=0.00125$, (b) $\Gamma/\sigma=0.0028125$, and (c) $\Gamma/\sigma=0.005$. Each graph shows results for code distances $d=3,5,7,9$, with $20\, 000$ simulation shots per data point.}
    \label{fig:1abc}

\end{figure}

We first fix the dimensionless coupling strength $\Gamma/\sigma$ and vary the number of atoms $N$. As shown in \cref{fig:1abc}, the logical error rate increases with the number of atoms, with stronger coupling leading to a faster increase. For each coupling strength, we observe threshold behavior across the four code distances $d=3,5,7,9$. The total logical error rate includes both failures of the minimum-weight perfect matching decoder and residual logical Pauli errors ($X$, $Y$, or $Z$). These results indicate that, for a fixed coupling strength, there is a threshold number of atoms below which increasing the code distance suppresses logical errors.

\begin{figure}[ht!]
\centering

    \noindent\makebox[0.5\linewidth][l]{\large\bfseries (a)}
    \vspace{0.4em}

    \includegraphics[width=0.5\linewidth]{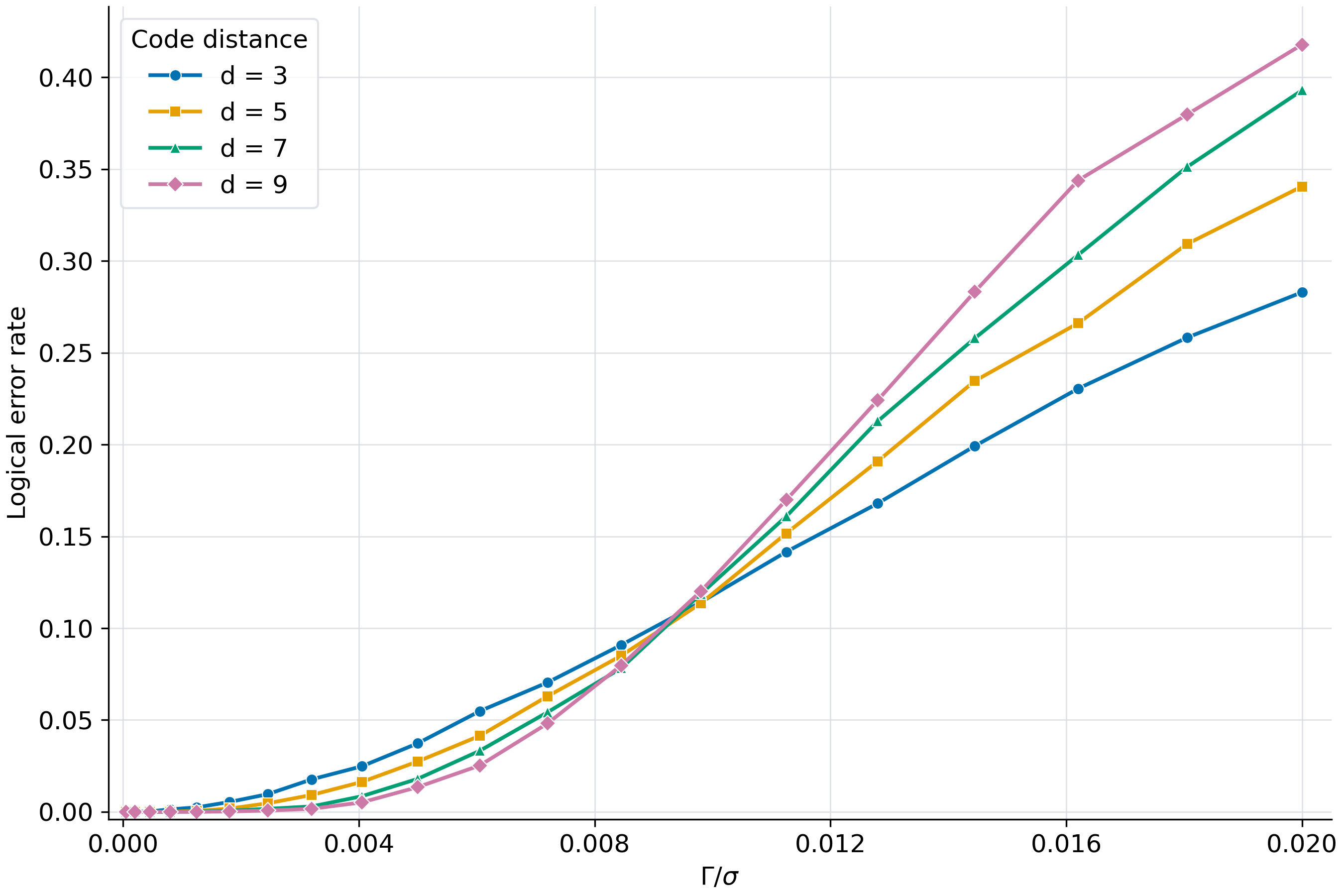}

    \vspace{0.4em}

    \noindent\makebox[0.5\linewidth][l]{\large\bfseries (b)}
    \vspace{0.4em}

    \includegraphics[width=0.5\linewidth]{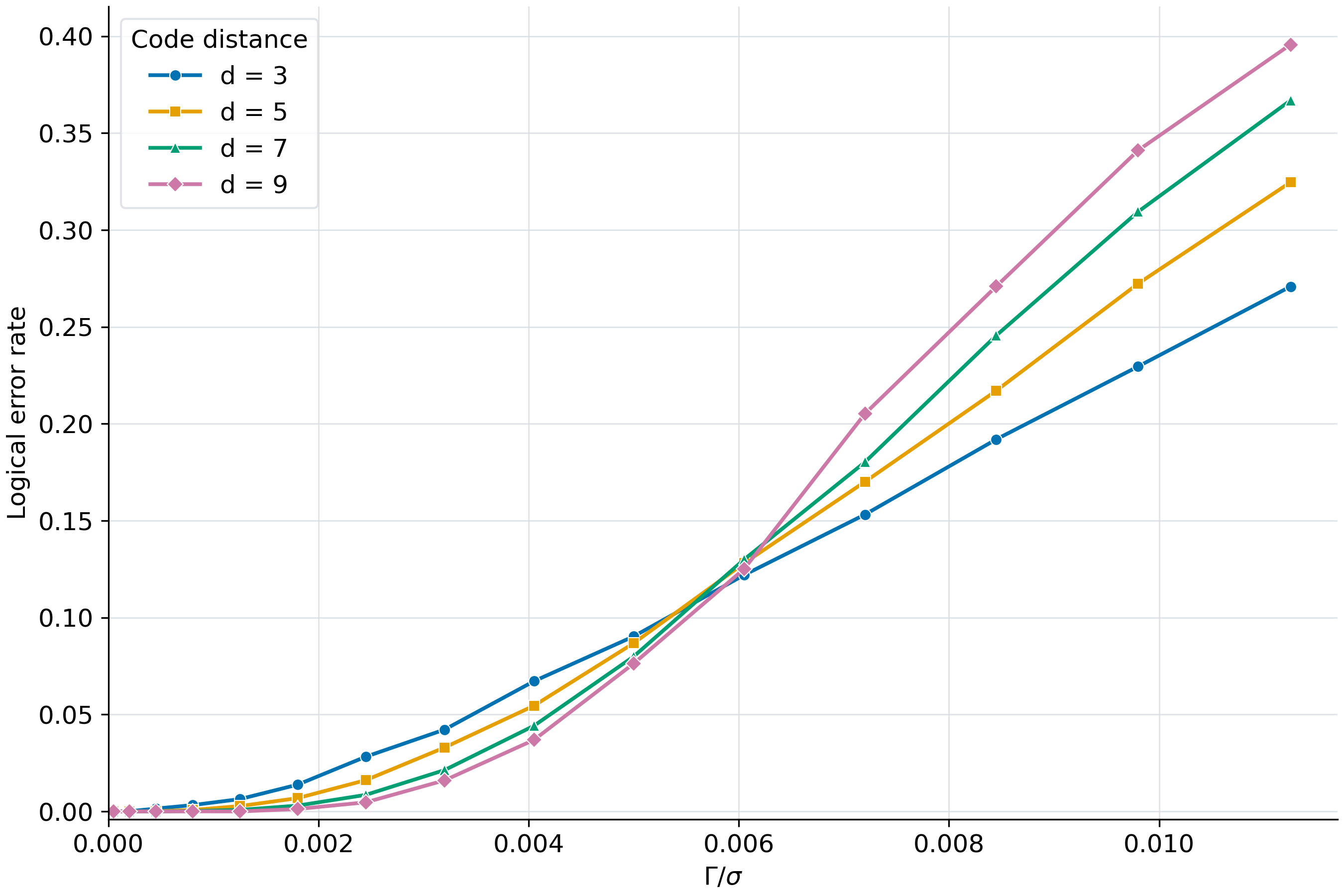}

    \vspace{0.4em}

    \noindent\makebox[0.5\linewidth][l]{\large\bfseries (c)}
    \vspace{0.4em}

    \includegraphics[width=0.5\linewidth]{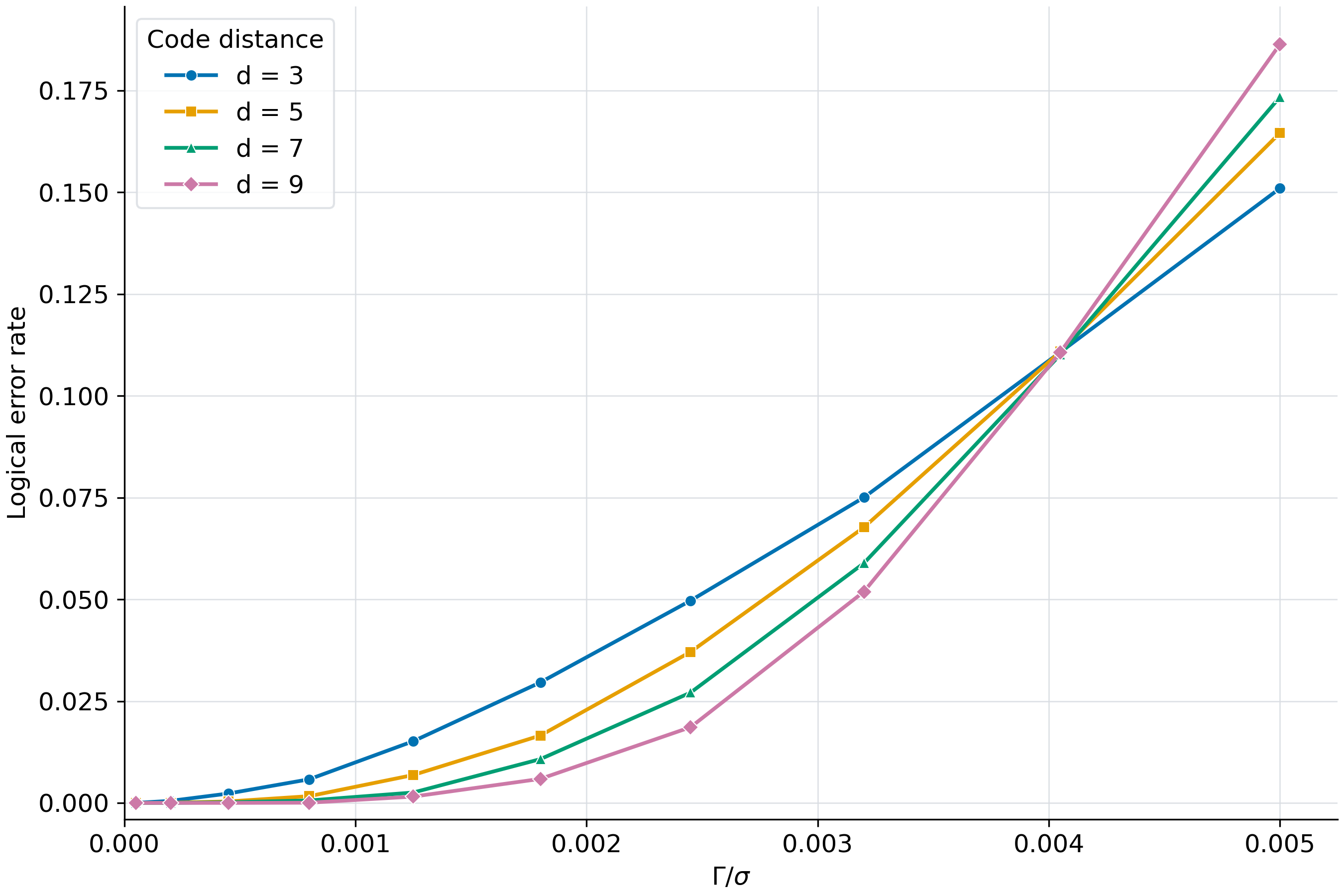}

    \caption{Logical error rate as a function of the dimensionless coupling strength $\Gamma/\sigma$ for fixed atom numbers (a) $N=3$, (b) $N=5$, and (c) $N=7$. Each graph shows results for code distances $d=3,5,7,9$, with $20\,000$ simulation shots per data point.}
    \label{fig:2abc}

\end{figure}

We next fix the number of atoms $N$ and vary the dimensionless coupling strength $\Gamma/\sigma$. As shown in \cref{fig:2abc}, we again observe threshold behavior, now as a function of the coupling strength. For each fixed atom number, increasing the code distance suppresses logical errors below a threshold coupling strength. Together, these results show that the error correction performance depends on both the number of atoms and the atom-waveguide coupling strength.

\subsection{Boundary Map}
\label{sec:BoundaryMap}
\begin{figure}[th!]
\centering
\includegraphics[width=\linewidth]{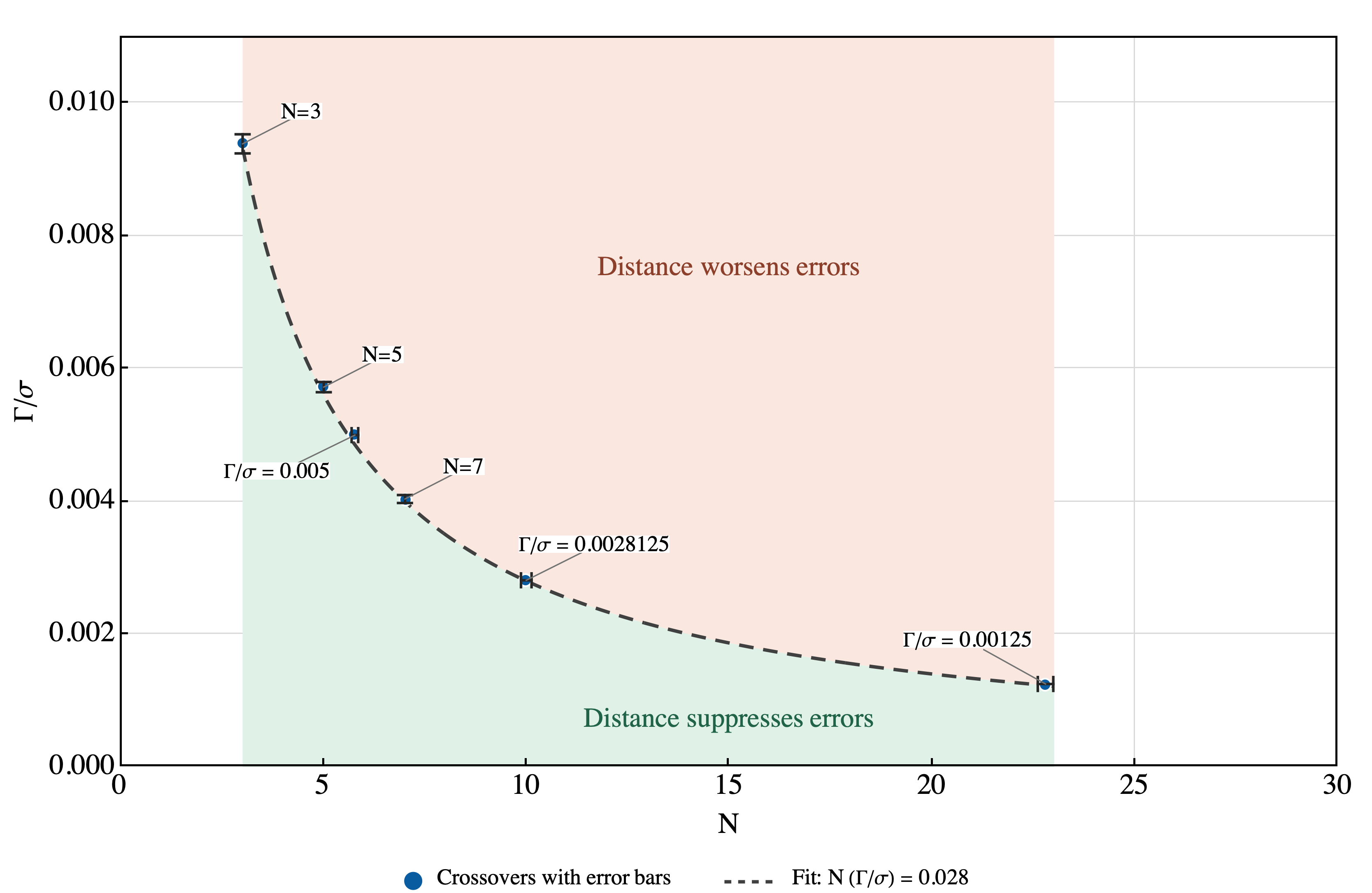}
\caption{Boundary map by plotting the crossing value for dimensionless coupling strength $\Gamma / \sigma$ together with the number of atoms $N$.\label{fig:BoundaryMap}}
\end{figure}

To characterize the threshold behavior, we construct a boundary map in \cref{fig:BoundaryMap} using six threshold points in \cref{fig:1abc,fig:2abc}. Details concerning about how to determine the threshold points as well as the standard deviation is provided in Appendix \ref{appendix:Scalinganalysisofthresholdpointsandstandarddeviation}. Plotting the number of atoms against the coupling strength, we find that the numerical relation
\begin{equation}
\frac{N\Gamma}{\sigma} = 0.028
\end{equation}
provides an excellent fit to these points.

We now show that this relation can be understood from the microscopic scattering model. For identical, lossless atoms with symmetric real coupling, the single-atom transfer matrix in \cref{eq:IndividualTransferMatrix} becomes
\begin{equation}
T(\omega)
=
\frac{1}{\Delta(\omega)+2i\Gamma }
\begin{pmatrix}
\Delta(\omega) & -i2\Gamma \\
-i2\Gamma  & \Delta(\omega)
\end{pmatrix}.
\end{equation}
The total transfer matrix can then be written as
\begin{equation}
U_N(\omega) = T^N(\omega) = \frac{1+s^N(\omega)}{2} I + \frac{s^N(\omega)-1}{2} X,
\end{equation}
where $s(\omega) =\exp\left(-2i\arctan{\left(\frac{2\Gamma}{\Delta(\omega)}\right)} \right)$. Using \cref{eq:PauliErrorProbability}, we obtain the Pauli error probabilities
\begin{equation}
p_Y = p_Z = 0,
\end{equation}
and
\begin{equation}
p_X = \int d\omega\,
\frac{e^{-\Delta(\omega)^2/\sigma^2}}
{\sigma\pi^{1/2}}
\sin^2
\left(
N\cdot
\arctan
\left(
\frac{2\Gamma}{\Delta(\omega)}
\right)
\right),
\end{equation}
where we have used the resonance condition $\omega_{center} = \Omega$.

In the limit $N\Gamma \ll \sigma$, the photon spectrum varies slowly over the relevant scattering bandwidth, allowing the exponential factor in the integral to be approximated by unity. We thus obtain
\begin{equation}
\begin{split}
p_X
&= 2\sqrt{\pi}\frac{N\Gamma}{\sigma} + O\left(\left(\frac{N\Gamma}{\sigma}\right)^2\right).
\end{split}
\end{equation}
Evaluating this expression at the six numerical threshold points and taking the mean gives a threshold Pauli $X$-error probability of approximately $p_X = 0.10$. This corresponds to $N\Gamma /\sigma = 0.028$, consistent with the numerical fit. Detailed analysis of the derivation and the error estimate are provided in the Appendix \ref{appendix:Boundarymapformulaanderrorestimate}.

\section{Quantum phase estimate algorithm}
\label{sec:QuantumPhaseEstimateAlgorithm}
Quantum phase estimation is a quantum algorithm that estimates the eigenphase of a unitary operator. Here, we use this algorithm to investigate how waveguide-induced errors affect algorithmic performance after quantum error correction. We assume ideal local quantum gates, and quantum information is transmitted through the waveguide between operations. The transmission noise is modeled using the logical Pauli error model obtained from the rotated planar code simulations. Finally, we use the identified error bias to modify the phase estimation circuit and reduce its sensitivity to the dominant logical error.
\subsection{Error model}
\label{sec:Quickintroductionanderrormodel2}
\begin{figure}[htbp]
\centering
\begin{quantikz}[row sep={0.9cm,between origins}, column sep=0.52cm]
\lstick{$\ket{0}_{q_0}$}
  & \gate{H} & \ctrl{3} & \qw       & \qw       & \gate[3]{F_{8}^{\dagger}} & \meter{} \rstick{$y_0$} \\
\lstick{$\ket{0}_{q_1}$}
  & \gate{H} & \qw      & \ctrl{2} & \qw       & \qw & \meter{}  \rstick{$y_1$} \\
\lstick{$\ket{0}_{q_2}$}
  & \gate{H} & \qw      & \qw       & \ctrl{1} & \qw & \meter{}  \rstick{$y_2$} \\
\lstick{$\ket{u}$}
  & \qw      & \gate{U} & \gate{U^2}& \gate{U^4}& \qw  & \rstick{$\ket{u}$}
\end{quantikz}
\caption{Quantum phase estimation with three ancilla qubits.}
\label{fig:QPE}
\end{figure}
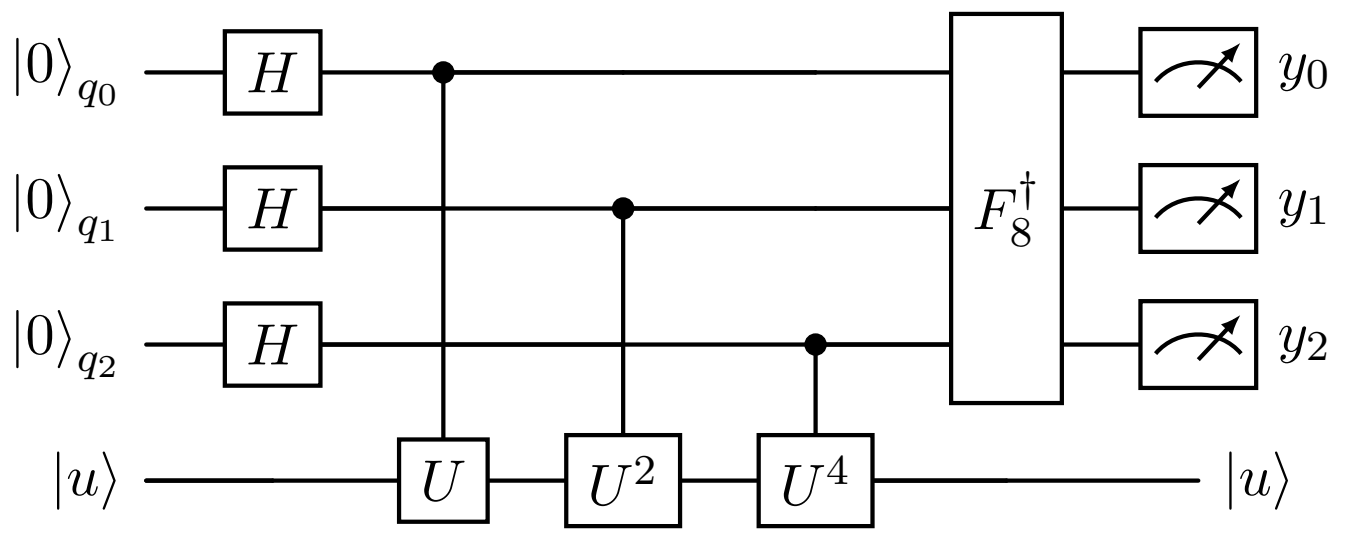

We first review the quantum phase estimation algorithm, and further details can be found in \cite{Cleve1998}. Consider a unitary operator $U$ with eigenstate $\ket{u}$
\begin{equation}
U\ket{u}=e^{2\pi i\theta}\ket{u},
\end{equation}
where $e^{2\pi i\theta}$ is the eigenvalue and $\theta\in[0,1)$ is unknown.
Quantum phase estimation estimates $\theta$ given the ability to prepare $\ket{u}$ and implement controlled powers of $U$. A circuit with three ancilla qubits is shown in  \cref{fig:QPE}.

We initialize the $m$ control qubits in $\ket{0}^{\otimes m}$ and the target register in $\ket{u}$. The initial state is therefore
\begin{equation}
\ket{0}^{\otimes m} \ket{u}. 
\end{equation}
Applying a Hadamard gate to each control qubit yields
\begin{equation}
\frac{1}{\sqrt{M}} \sum_{k_0,\cdots, k_{m-1} \in \{0,1\}} \ket{k_{m-1}\cdots k_0}\ket{n} = \frac{1}{\sqrt{M}} \sum_{k = 0}^{M-1} \ket{k}\ket{n}
\end{equation}
where $\ket{k} = \sum_{j = 0}^{m-1} k_j 2^j$ and $M = 2^m$. We then apply a controlled-$U^{2^j}$ gate, with the $j$th control qubit acting on the target register, for $j=0, \cdots, m-1$. The resulting state is
\begin{equation}
\frac{1}{\sqrt{M}} \sum_{k =0}^{M-1} e^{2\pi i k \theta} \ket{k}\ket{u}. 
\end{equation}
Applying the inverse quantum Fourier transform $F_M^\dagger$ to the control register yields
\begin{equation}
\sum_{x = 0}^{M-1} A_x\ket{x} \ket{u},
\end{equation}
where $A_x = \frac{1}{M} \sum_{k = 0}^{M-1} e^{2\pi i k (\theta - x/M)}$ is the probability amplitude for the state $\ket{x}\ket{u}$. Here, the quantum Fourier transform is defined as
\begin{equation}
F_M\ket{x} = \frac{1}{\sqrt{M}}\sum_{k = 0}^{M-1} e^{2\pi i x k /M} \ket{k}.
\end{equation}
If $\theta = \frac{b}{M}$ for some $b = 0, \cdots, M-1$, the probability amplitude of the state $\ket{x}\ket{u}$ reduces to
\begin{equation}
A_x = \frac{1}{M} \sum_{k = 0}^{M-1} e^{2\pi i k (b- x)/M}.
\end{equation}
For $x  = b$, we obtain $A_b = 1$, corresponding to perfect phase estimation. Away from this exactly representable case, we define the estimation error as $\varepsilon = \theta - x/M$, giving
\begin{equation}
A_x = \frac{1}{M} \sum_{k = 0}^{M-1} e^{2\pi i k \varepsilon}.
\end{equation}
The corresponding measurement probability is
\begin{equation}
P(y) = \abs{A_x}^2 = \frac{1}{M^2} \frac{\sin^2(M\varepsilon\pi)}{\sin^2(\varepsilon\pi)}.
\end{equation}

We now incorporate waveguide-induced errors into the phase estimation circuit. We assume that all quantum gates, including the inverse quantum Fourier transform, and measurements are performed locally and ideally. We further assume that quantum information is transmitted through the waveguide between spatial locations. The transmission noise is modeled by the logical Pauli errors obtained from the rotated planar code simulations. In particular, each controlled-$U^{2^j}$ gate is treated as a single local operation, rather than as $2^j$ separate waveguide transmissions. Although we do not consider Kitaev's phase estimation method \cite{Kitaev1995}, the same framework could be applied to that implementation.

Our quantum phase estimation simulation uses the logical Pauli-error probabilities obtained from the rotated planar code simulations at fixed coupling strength and code distance $d=9$. These errors are applied at the modeled transmission steps between local operations. We then evaluate the phase estimation performance as a function of the number of atoms, restricting the analysis to the regime below the error correction threshold. The simulation evolves the quantum state vector exactly.

\subsection{Results}
\label{sec:results}
We first investigate the performance of quantum phase estimation for $m=4,6,8$ ancilla qubits, using a target phase of $\theta=7/8=0.875$ and several coupling strengths. For each parameter setting, we average the measured phase over $20\, 000$ shots and restrict the number of atoms to values below the corresponding error correction threshold.

\begin{figure}[t]
\centering
\includegraphics[width=0.8\linewidth]{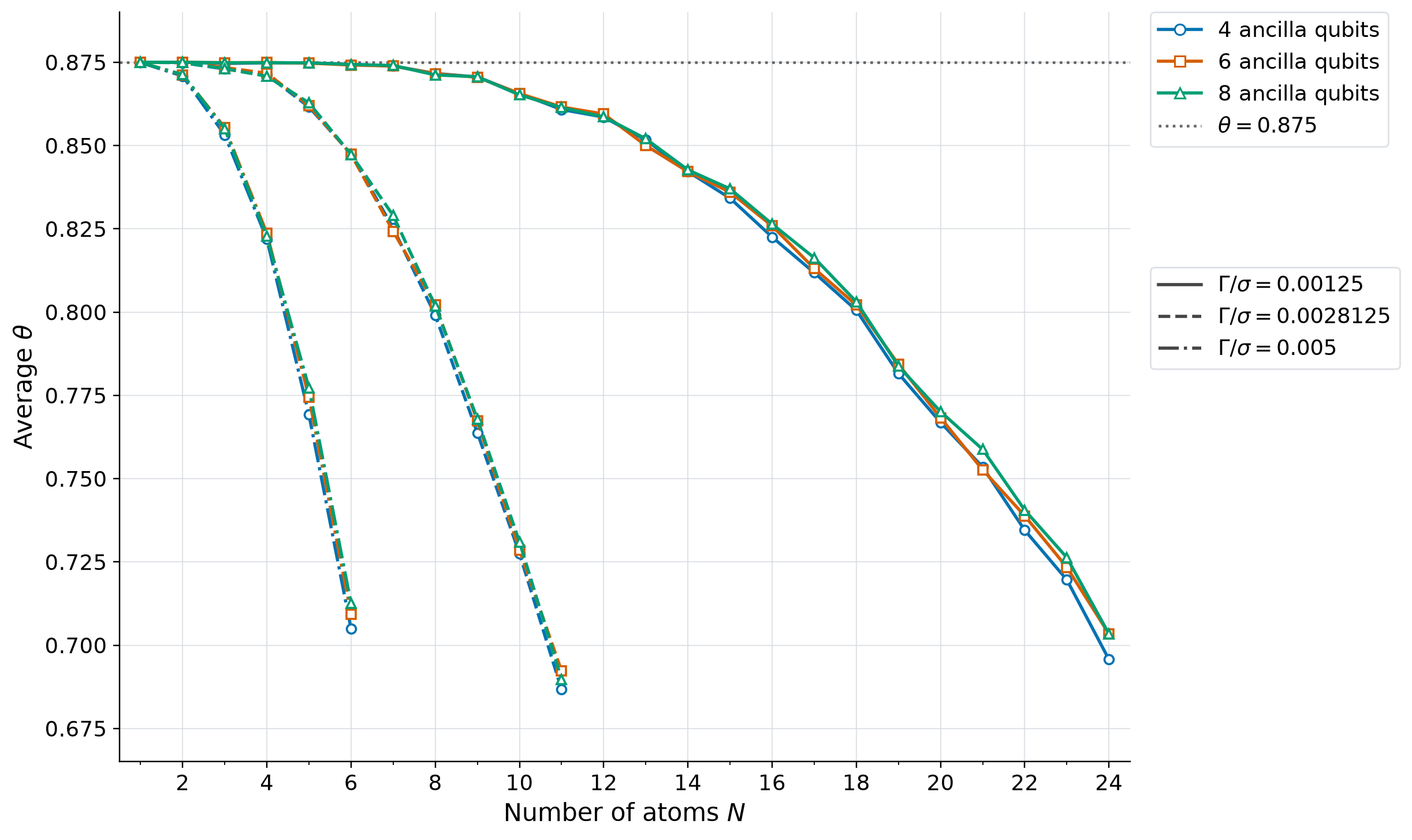}
\caption{Numerical simulation of quantum phase estimation with true phase $7/8 = 0.875$, showing the average measured phase as a function of the number of atoms.\label{fig:StandardQPE}}
\end{figure}

\begin{figure}[t]
\centering
\includegraphics[width=0.8\linewidth]{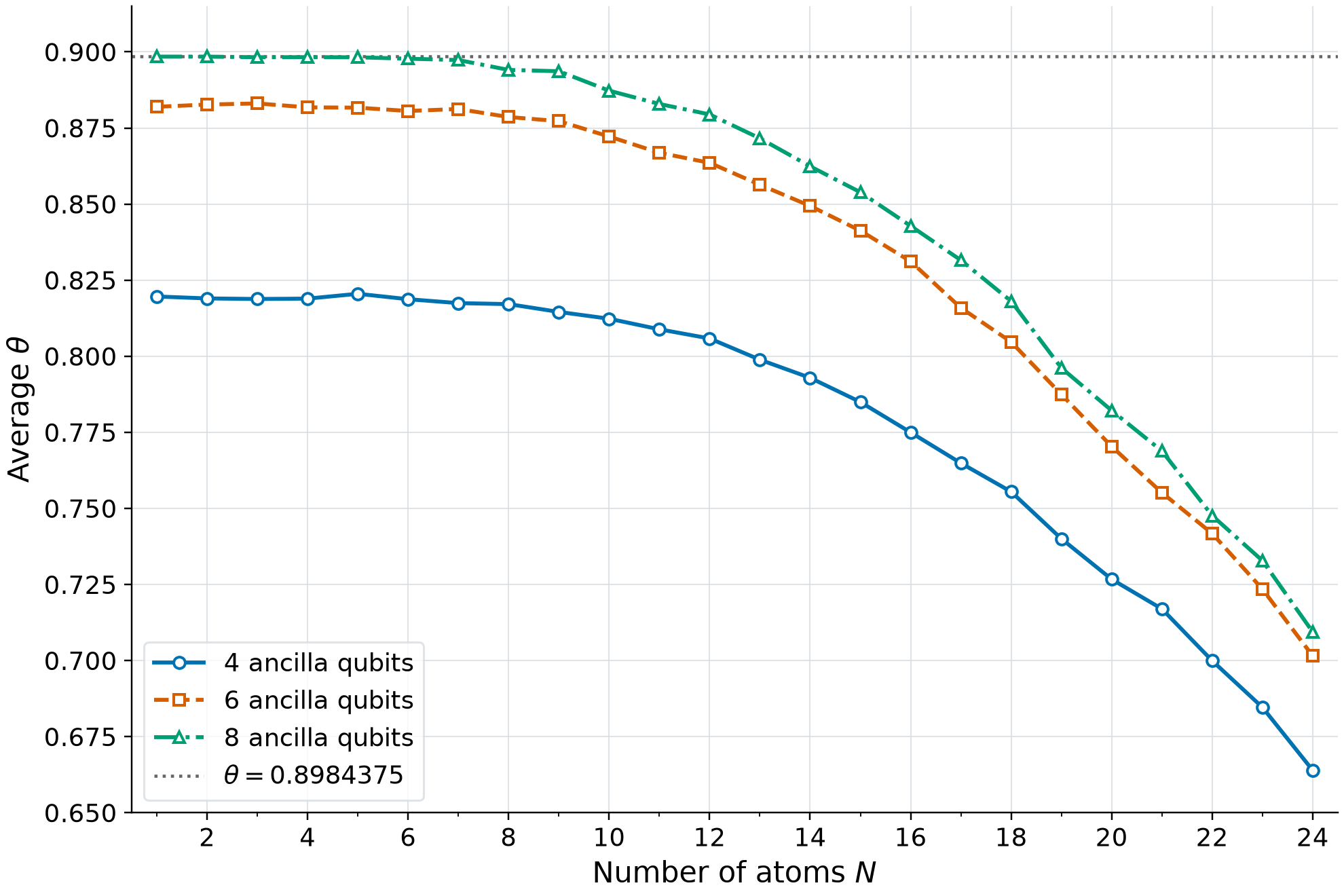}
\caption{Numerical simulation of quantum phase estimation with true phase $115/128 = 0.8984375$, showing the average measured phase as a function of the number of atoms.\label{fig:AncillaImprovePrecision}}
\end{figure}

As shown in \cref{fig:StandardQPE}, the average estimated phase deviates further from the target as the number of atoms $N$ increases. Stronger coupling $\Gamma/\sigma$ leads to larger deviations, while weaker coupling preserves the estimation accuracy over a wider range of $N$. The number of ancilla qubits has little effect in this case because the target phase $7/8$ can be represented exactly for all three values of $m$. To examine the effect of the ancilla register size, we next consider $\theta=115/128$, as shown in \cref{fig:AncillaImprovePrecision}. Unlike the previous case, this phase can be represented exactly with $m=8$ ancilla qubits, but not with $m=4$ or $6$. The dependence on the number of ancilla qubits is therefore more pronounced, illustrating the benefit of increased phase resolution.


\subsection{One improvement}
\label{sec:OneImprovement}
The boundary-map analysis in \cref{sec:BoundaryMap} shows that the effective logical noise is dominated by $X$ errors. This suggests modifying quantum phase estimation so that the ancilla qubits are encoded and measured in the Hadamard basis, where $X$ acts diagonally. Starting from the standard circuit, this basis change is implemented by conjugating the ancilla operations by Hadamard gates. Although one might then prepare each ancilla in $\ket{+}$, this would be immediately followed by a Hadamard gate, making the first step redundant. It is therefore equivalent to initialize the ancilla qubits in $\ket{0}$.

\begin{figure}[htbp]
\centering
\begin{quantikz}[row sep={0.9cm,between origins}, column sep=0.52cm]
\lstick{$\ket{0}_{q_0}$}
   & \xctrl{3} & \qw       & \qw       & \gate[3]{H^{\otimes 3}F_{8}^{\dagger}H^{\otimes 3}} & \xmeter & \wireoverride{n} \rstick{$y_0$} \\
\lstick{$\ket{0}_{q_1}$}
   & \qw      & \xctrl{2} & \qw       & \qw & \xmeter & \wireoverride{n} \rstick{$y_1$} \\
\lstick{$\ket{0}_{q_2}$}
   & \qw      & \qw       & \xctrl{1} & \qw & \xmeter & \wireoverride{n}  \rstick{$y_2$} \\
\lstick{$\ket{u}$}
      & \gate{U} & \gate{U^2}& \gate{U^4}& \qw  & &\rstick{$\ket{u}$}
\end{quantikz}
\caption{Modified quantum phase estimation with three ancilla qubits using Hadamard basis.}
\label{fig:ModifiedQPEAlgorithm}
\end{figure}
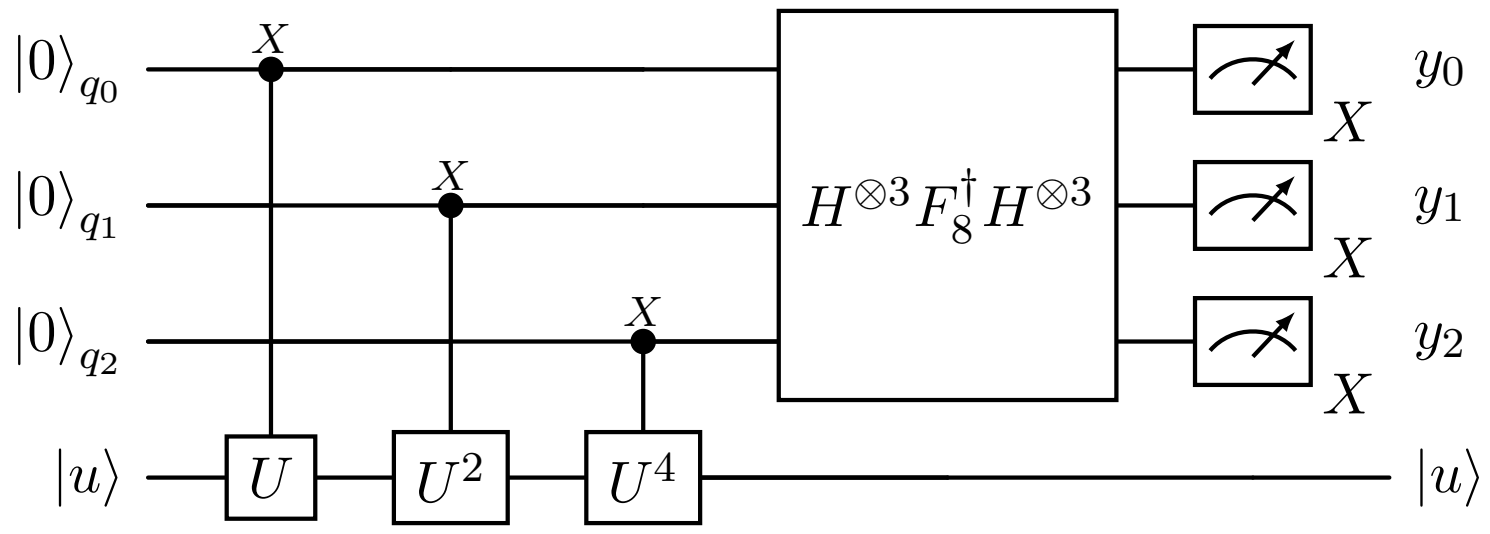
In \cref{fig:ModifiedQPEAlgorithm},  we present the circuit for the modified algorithm, where we define
\begin{equation}
\begin{quantikz}[row sep={0.9cm,between origins}, column sep=0.52cm]
& \xctrl{1} & \qw \\
  & \gate{U^{2^j}} & \qw
\end{quantikz}
\;=\;
\begin{quantikz}[row sep={0.9cm,between origins}, column sep=0.52cm]
& \gate{H} & \ctrl{1} & \gate{H} & \qw \\
  & \qw & \gate{U^{2^j}} & \qw & \qw
\end{quantikz}
\end{equation}
and 
\begin{equation}
\begin{quantikz}
\xmeter
\end{quantikz}
\end{equation}
denotes measurement in the $X$ basis. In this basis, the controlled-$U^{2^j}$ gate is replaced by
\begin{equation}
\ket{+}\bra{+}_{j}\otimes I + \ket{-}\bra{-}_{j}\otimes U^{2j} = H_j(\ket{0}\bra{0}_{j}\otimes I + \ket{1}\bra{1}_{j}\otimes U^{2j})H_j,
\end{equation}
which is the controlled-$U^{2^j}$ operation expressed in the rotated basis. Similarly, we replace the inverse quantum Fourier transform $F_M^\dagger$ by $H^{\otimes m}F_M^\dagger H^{\otimes m}$, while the final measurement is performed directly in the $X$ basis. The $H$ gate itself is invariant under this basis transformation.

\begin{figure}[th!]
\centering
\includegraphics[width=0.8\linewidth]{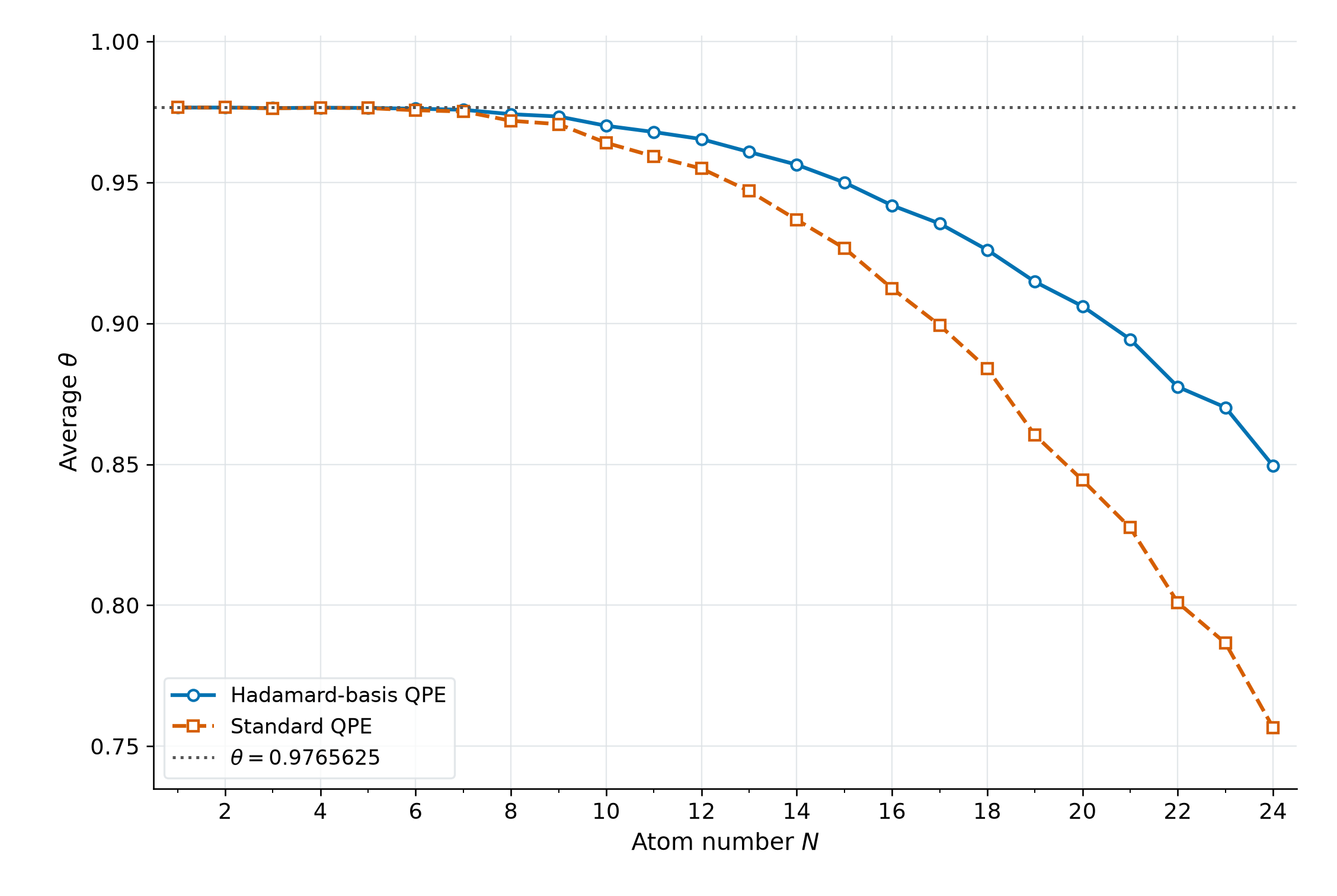}
\caption{Comparison of the standard and Hadamard-basis implementations of quantum phase estimation for $\Gamma/\sigma=0.00125$. The Hadamard-basis implementation shows improved performance under the $X$-biased error model.\label{fig:ModifiedQPEComparison}}
\end{figure}

We adopt the same assumptions as in the previous subsection: all gates are local and ideal, while transmission errors are inserted after each gate operation. We then repeat the simulation for the modified circuit at the same coupling strength, $\Gamma/\sigma=0.00125$. As shown in \cref{fig:ModifiedQPEComparison}, the Hadamard-basis implementation yields a clear improvement under this error model.

Although the standard and modified circuits are equivalent in the absence of noise, they behave differently once errors are inserted at specific circuit locations. This difference can be understood by tracking how the dominant $X$ errors transform under the basis change. For an $X_j$ error acting on the $j$th ancilla between the $X$-basis controlled-$U^{2^j}$ gate and the Hadamard-basis inverse quantum Fourier transform, we have
\begin{equation}
\begin{split}
\begin{quantikz}[row sep={0.7cm,between origins}, column sep=0.3cm]
\lstick{$q_j$}
 & \gate{H} & \ctrl{2} & \gate{H} & \gate{X}
 & \gate[2]{H^{\otimes m}}
 & \gate[2]{F_{2^m}^{\dagger}}
 & \gate[2]{H^{\otimes m}} & \qw \\
\lstick{$\bq_{\ne j}$}
 & \qw & \qw & \qw & \qw
 & \qw & \qw & \qw & \qw \\
\lstick{$u$}
 & \qw & \gate{U^{2^j}} & \qw & \qw
 & \qw & \qw & \qw & \qw
\end{quantikz}\\
{}=\quad
\begin{quantikz}[row sep={0.7cm,between origins}, column sep=0.3cm]
\lstick{$q_j$}
 & \gate{H} & \ctrl{2} & \gate{Z}
 & \gate[2]{H^{\otimes m}H_j}
 & \gate[2]{F_{2^m}^{\dagger}}
 & \gate[2]{H^{\otimes m}} & \qw \\
\lstick{$\bq_{\ne j}$}
 & \qw & \qw & \qw
 & \qw & \qw & \qw & \qw \\
\lstick{$u$}
 & \qw & \gate{U^{2^j}} & \qw
 & \qw & \qw & \qw & \qw
\end{quantikz}
\end{split}
\end{equation}
Here, $\bq_{\ne j}$ denotes the set of ancilla qubits other than the $j$th qubit. Under this transformation, the $X_j$ error becomes a $Z_j$ error. Such an error can still affect the computation because $Z_j$ does not commute with $F_{2^m}^\dagger$.

By contrast, for an $X_j$ error occurring between the Hadamard-basis inverse quantum Fourier transform and the final $X$-basis measurement, we have
\begin{equation}
\begin{split}
&\begin{quantikz}[row sep={0.7cm,between origins}, column sep=0.3cm]
\lstick{$q_j$}
 & \gate[2]{H^{\otimes m}}
 & \gate[2]{F_{2^m}^{\dagger}}
 & \gate[2]{H^{\otimes m}}
 & \gate{X} & \xmeter & \wireoverride{n} \rstick{$y_j$} \\
\lstick{$\bq_{\ne j}$}
 & \qw & \qw & \qw & \qw & \qw & \qw
\end{quantikz}\\
{}=\quad
&\begin{quantikz}[row sep={0.7cm,between origins}, column sep=0.3cm]
\lstick{$q_j$}
 & \gate[2]{H^{\otimes m}}
 & \gate[2]{F_{2^m}^{\dagger}}
 & \gate[2]{H^{\otimes m}}
 & \gate{H} & \gate{Z} & \meter{} & \wireoverride{n}  \rstick{$y_j$} \\
\lstick{$\bq_{\ne j}$}
 & \qw & \qw & \qw & \qw & \qw & \qw & \qw
\end{quantikz}\\
{}=\quad
&\begin{quantikz}[row sep={0.7cm,between origins}, column sep=0.3cm]
\lstick{$q_j$}
 & \gate[2]{H^{\otimes m}}
 & \gate[2]{F_{2^m}^{\dagger}}
 & \gate[2]{H^{\otimes m}}
 & \gate{H} & \meter{} & \wireoverride{n} \rstick{$y_j$} \\
\lstick{$\bq_{\ne j}$}
 & \qw & \qw & \qw & \qw & \qw & \qw
\end{quantikz}
\end{split}
\end{equation}
In this case, the dominant error does not change the final measurement statistics. The Hadamard-basis implementation therefore contains fewer locations at which the dominant $X$ errors are harmful, providing a simple explanation for the observed improvement.

\section{Discussion}
\label{sec:discussion}
The main contribution of this work is a first-principles framework connecting microscopic waveguide quantum optics to quantum error correction and algorithmic performance. Starting from the Hamiltonian for single-photon scattering by an ensemble of two-level atoms, we derive the polarization-dependent scattering matrix and construct an effective quantum channel by tracing out the frequency degree of freedom.

Using the resulting effective noise model, we study quantum error correction with the rotated planar code. For identical, lossless atoms with symmetric real coupling, the numerical simulations show a threshold behavior when the number of atoms $N$ or the coupling strength $\Gamma$ is varied. The threshold boundary is approximately described by
\begin{equation}
\frac{N\Gamma}{\sigma}= 0.028,
\end{equation}
where $\sigma$ is the bandwidth. An asymptotic analysis of the microscopic scattering model in the regime $N\Gamma\ll\sigma$ explains this scaling. For a photon spectrum centered at the atomic resonance, the effective noise contains only $X$ errors, with logical error probabilities $p_Y=p_Z=0$. Evaluating the asymptotic $X$-error probability at the numerical threshold points gives average $p_X=0.10$, consistent with the fitted boundary. These results connect the observed error correction threshold behavior to the microscopic scattering model.

We then incorporate the resulting logical errors into quantum phase estimation, which relates algorithmic performance directly to the microscopic waveguide parameters. As the atom-induced noise increases, the phase estimation accuracy degrades. In particular, the strong $X$ bias guides the circuit design. We use this bias to construct a Hadamard-basis implementation of phase estimation that is less sensitive to the dominant error. Although the standard and modified circuits are equivalent without noise, they respond differently to errors introduced between operations. In particular, $X$ errors immediately before the final $X$-basis measurement leave the measurement statistics unchanged. The observed improvement therefore illustrates how microscopic knowledge of the noise can guide algorithm design beyond the choice of error correction code.

The framework in this study can be extended to more realistic waveguide systems and other quantum-information platforms, such as circuit QED, neutral-atom, and free-space photonic systems. Within the waveguide model, photon loss could be included through atomic decay or an open-system model. Other potential directions are to study disorder in atomic transition frequencies, coupling strengths, number of atoms, and positions and to consider bidirectional waveguides which would include effects like reflection and backscattering. One important extension is to include microscopic descriptions of imperfections in state preparation, one-qubit and two-qubit gates, ancilla control, and measurement for a complete end-to-end analysis. The strong $X$ bias identified in this study also motivates exploring error correction codes tailored to biased noise, such as the XZZX surface code \cite{BonillaAtaides2021} and other bias-aware surface codes \cite{Tuckett2019,Aliferis2008}.

Such extensions could also require a detailed treatment of logical operations. In the surface code, the logical operators $X_L$ and $Z_L$ are strings of physical Pauli operators connecting opposite boundaries. Usually, logical Pauli corrections or operations can be tracked on classical software and updated in the Pauli frame \cite{Fowler2012}. Logical CNOT gates and magic-state injection can be implemented through lattice surgery by merging and splitting, while logical Hadamard gates can be realized using transversal physical Hadamard operations combined with lattice surgery to maintain orientation  \cite{Horsman2012, Litinski2019}. 

More broadly, this work suggests that information obtained from a microscopic noise model can inform not only error correction but also the design and implementation of quantum algorithms. In particular, microscopic noise models could be used to compare equivalent circuit implementations. Such an approach is especially relevant to early fault-tolerant quantum computing, where error correction resources are limited. Ultimately, a microscopic noise model could provide a joint optimization of physical encoding, error correction code, code distance, and algorithm implementation, with the goal of achieving a target accuracy using fewer physical resources.

\begin{acknowledgments}
I would like to thank Professor John C. Schotland for his invaluable guidance and support, and for providing the model used in this work. I would also like to thank Profressor Steven M. Girvin and Professor Imran M. Mirza for their helpful discussions and suggestion.
\end{acknowledgments}

\appendix
\section{Arbitrary qubit rotation}
\label{appendix:Arbitraryqubitrotation}
In this section, we will show that the transfer matrix $T$ is an arbitrary qubit rotation
\begin{equation}
T = e^{-i\theta} R_{\hat{n}}(2\theta),
\end{equation}
where the rotation angle is
\begin{equation}
\theta=\arctan\left(\frac{\Gamma_{+}+\Gamma_{-}}{\omega-\Omega}\right) 
\end{equation}
and rotation axis is
\begin{equation}
\hat{n} = \frac{1}{|V_{+}|^2+|V_{-}|^2}
\begin{bmatrix}
2\operatorname{Re}\bigl(V_{+}^{*}V_{-}\bigr)\\
2\operatorname{Im}\bigl(V_{+}^{*}V_{-}\bigr)\\
|V_{+}|^2-|V_{-}|^2
\end{bmatrix}.
\end{equation}
Consider the eigenvector of $T$
\begin{equation}
\ket{B}=\frac{V_{+}\ket{0}+V_{-}\ket{1}}
{\sqrt{|V_{+}|^2+|V_{-}|^2}},
\qquad
\ket{D}=\frac{V_{-}^{*}\ket{0}-V_{+}^{*}\ket{1}}
{\sqrt{|V_{+}|^2+|V_{-}|^2}},
\end{equation}
with eigenvalues $\frac{\Delta-i\bigl(\Gamma_{+}+\Gamma_{-}\bigr)}
{\Delta+i\bigl(\Gamma_{+}+\Gamma_{-}\bigr)}$ and $1$ respectively.
One can verify orthogonality and normalization
\begin{equation}
\langle B|D\rangle = 0, \quad \langle B|B\rangle =\langle D|D\rangle = 1.
\end{equation}
Then, starting from \cref{eq:IndividualTransferMatrix}, direct computation gives
\begin{equation}
\begin{split}
T&=\ket{D}\bra{D}
+\frac{\Delta-i\bigl(\Gamma_{+}+\Gamma_{-}\bigr)}
{\Delta+i\bigl(\Gamma_{+}+\Gamma_{-}\bigr)}\ket{B}\bra{B}\\
&=\ket{D}\bra{D}+e^{-2i\theta}\ket{B}\bra{B}\\
&= e^{-i\theta}\left[\cos\theta I+i\sin\theta
\left(\ket{D}\bra{D}-\ket{B}\bra{B}\right)\right]\\
&=e^{-i\theta}\left[\cos\theta I+i\sin\theta
\left(\frac{I-\hat n\cdot\hat\sigma}{2}-\frac{I+\hat n\cdot\hat\sigma}{2}\right)\right]\\
&=e^{-i\theta}\left[\cos\theta I-i\sin\theta\hat n\cdot\hat\sigma\right]
=e^{-i\theta}e^{-i\theta\hat n\cdot\hat\sigma}
=e^{-i\theta}R_{\hat n}(2\theta),
\end{split}
\end{equation}
where $\hat \sigma = (\sigma_x, \sigma_y, \sigma_z)$ is the Pauli vector and $I$ is the identity matrix.

\section{Pauli-twirling channel}
\label{appendix:Pauli-twirlingchannel}
We define the Pauli-twirling channel for channel $\mathcal{N}$ to be 
\begin{equation}
\mathcal N_{PT}[P]=\frac{1}{4}\sum_{P\in\{I,X,Y,Z\}}
P\mathcal N(P\rho P)P.
\end{equation}
Direct evaluation gives
\begin{equation}
\mathcal N_{PT}[P] =\int d\omega\,\frac{1}{4}\sum_{P\in\{I,X,Y,Z\}}
P U_N(\omega)P\rho P U_N^{\dagger}(\omega)P|f(\omega)|^2.
\end{equation}
Each $U_N(\omega)$ can be written as
\begin{equation}
U_N(\omega) = \sum_P u_P(\omega)P,
\end{equation}
for some $u_P(\omega)$. To obtain such $u_P(\omega)$, we perform Hilbert-Schmidt inner product which is defined as 
\begin{equation}
\langle A,B\rangle_{HS}=\operatorname{Tr}(A^{\dagger}B)
\end{equation}
and get
\begin{equation}
u_P(\alpha)=\frac{1}{2}\operatorname{Tr}\bigl(PU_N(\alpha)\bigr).
\end{equation}
Then we calculate explicitly and find
\begin{equation}
\frac{1}{4}\sum_{P\in\{I,X,Y,Z\}}
P U_N(\alpha)P\rho P U_N^{\dagger}(\alpha)P = \sum_P|u_P(\alpha)|^2P\rho P. 
\end{equation}
Therefore, we have
\begin{equation}
\mathcal N_{PT}[\rho]=\int d\alpha\,|f(\alpha)|^2 \sum_P|u_P(\alpha)|^2P\rho P.
\end{equation}

Now, we will show that under lossless, symmetric coupling identical atoms at resonance, the Pauli Twirl channel is equal to the original channel (i.e. $\mathcal{N}_{PT} = \mathcal{N}$). We start by direct computation of the total transfer matrix which gives
\begin{equation}
U_N(\omega) = T^N = \frac{1+s^N}{2} I + \frac{s^N-1}{2} X
\end{equation}
where $s = e^{-2i\theta}$ and $\theta = \arctan{\frac{2\Gamma}{\Delta}}$, $\Delta = \omega - \Omega$ where $\Omega$ is the atomic frequency. Then, we calculate 
\begin{equation}
\begin{split}
U_N(\omega)\rho U_N^\dagger(\omega) &= \left(\frac{1+s^N}{2} I + \frac{s^N-1}{2} X\right)\rho\left(\frac{1+\bar{s}^N}{2} I + \frac{\bar{s}^N-1}{2} X\right)\\
&= \abs{\frac{1+s^N}{2}}^2\rho + \frac{1+s^N}{2}\frac{\bar{s}^N-1}{2} \rho X\\
&+ \frac{1+\bar{s}^N}{2}\frac{s^N-1}{2} X\rho + \abs{\frac{s^N-1}{2}}^2 X\rho X.
\end{split}
\end{equation}
Then, when we evaluate the integral, we have
\begin{equation}
\mathcal{N}[\rho] = p_I \rho + p_{IX}\rho X + p_{XI} X\rho + p_X X\rho X,
\end{equation}
where 
\begin{equation}
p_X = \int d\omega \abs{f(\omega)}^2 \abs{\frac{s^N-1}{2}}^2, \ p_I =\int d\omega \abs{f(\omega)}^2 \abs{\frac{s^N+1}{2}}^2
\end{equation}
which is exactly the same Pauli-Twirl Probability, but with off diagonal term
\begin{equation}
p_{XI} = \int d\omega \abs{f(\omega)}^2 \frac{1+s^N}{2}\frac{\bar{s}^N-1}{2} = \bar{p}_{IX}.
\end{equation}
Now, we will show that these off-diagonal term will be zero when we assume resonance $\omega_{center} = \Omega$. At resonance, we can make a change of variable 
\begin{equation}
p_{XI} = \int d\Delta \abs{f(\Delta)}^2 \frac{1+s^N(\Delta)}{2}\frac{\bar{s}^N(\Delta)-1}{2}.
\end{equation}
Then, we see that 
\begin{equation}
\frac{1+s^N(-\Delta)}{2}\frac{\bar{s}^N(-\Delta)-1}{2} = -\frac{1+s^N(\Delta)}{2}\frac{\bar{s}^N(\Delta)-1}{2}
\end{equation}
which is an odd function and $\abs{f(\Delta)}^2$ is even. Thus, the integral will evaluate to zero $P_{XI} = 0 = P_{IX}$ which we prove that in this case the Pauli-twirl channel is exactly the original channel.

\section{Scaling analysis of threshold points and standard deviation}
\label{appendix:Scalinganalysisofthresholdpointsandstandarddeviation}
We perform a scaling analysis to accurately determine the critical point. For each threshold point, we follow \cite{WANG2003} and fit the logical error rate using the following model:
\begin{equation}
p_L = A + B (p_X(N,\Gamma/\sigma)- p_c) d^{1/\kappa} + C(p_X(N,\Gamma/\sigma)- p_c)^2 d^{2/\kappa},
\end{equation}
where $p_L$ is the logical error rate, $p_X(N,\Gamma/\sigma)$ is the $X$ error probability, $p_c$ is the critical threshold probability, and $A, B, C, \kappa$ are constants determined separately for each graph in  \cref{fig:1abc,fig:2abc}.

We assume the uncertainty in the observed logical error rate can be described by the binomial distribution. Accordingly, the standard deviation is calculated based on 
\begin{equation}
\sigma = \sqrt{\frac{p_{L,c}(1-p_{L,c})}{D}}
\end{equation}
where $p_{L,c} = p_L(p_X = p_c) = A$ and $D = 20000$ is the total number of simulation per data point. 
\section{Boundary map formula and error estimate}
\label{appendix:Boundarymapformulaanderrorestimate}
We are trying to analyze
\begin{equation}
p_X=\int d\omega\,
\frac{e^{-\Delta(\omega)^2/\sigma^2}}
{\sigma\pi^{1/2}}
\sin^2
\left(
N\cdot
\arctan
\left(
\frac{2\Gamma}{\Delta(\omega)}
\right)
\right),
\end{equation}
for the limit $N\Gamma\ll\sigma$. We define the approximation to be
\begin{equation}
p_X' = \int_{-\infty}^{\infty}
d\Delta\,
\frac{1}{\sqrt{\pi}\sigma}
\sin^2
\left(
N\cdot
\arctan
\left(
\frac{2\Gamma}{\Delta}
\right)
\right).
\end{equation}
We make a change of variable defining $\varphi = N \frac{2\Gamma}{\Delta}$ and we have
\begin{equation}
p_X' = \frac{4\Gamma}{\sqrt{\pi}\sigma N}
\int_{0}^{\frac{N\pi}{2}}
\frac{\sin^2(\varphi)}
{\sin^2(\varphi/N)}
\,d\varphi.
\end{equation}
By utilizing the identity which is true for positive integer $N$
\begin{equation}
\begin{split}
\frac{\sin^2(\varphi)}
{\sin^2(\varphi/N)} &= \sum_{j=0}^{N-1}
\sum_{k=0}^{N-1}
e^{2i(j-k)\varphi/N}\\
&= N
+
2\sum_{m=1}^{N-1}
(N-m)\cos\left(\frac{2m\varphi}{N}\right).
\end{split}
\end{equation}
After taking the integral from $0$ to $N\pi/2$, we see the second part of the intergrand vanish and gives us
\begin{equation}
p_X' = \frac{2\sqrt{\pi} N\Gamma}{\sigma}.
\end{equation}
For error estimate. We directly evaluate the difference
\begin{equation}
p_X' - p_X = \int d\omega\,
\frac{
1-e^{-\Delta^2/\sigma^2}
}{
\sigma\pi^{1/2}
}
\sin^2
\left(
N\cdot
\arctan
\left(
\frac{2\Gamma}{\Delta(\omega)}
\right)
\right)
\end{equation}
which is a positive number. Let $\theta = \arctan
\left(
\frac{2\Gamma}{\Delta(\omega)}\right)$. We know that for positive integer $N$, we have the identity
\begin{equation}
|\sin(N\theta)| \leq N|\sin(\theta)|.
\end{equation}
Thus, we have that 
\begin{equation}
\begin{split}
\sin^2(N\theta) &< N^2\sin^2(\theta) \\
&= N^2 \frac{\tan^2\theta}{1+\tan^2\theta} \\
&=N^2 \frac{\left(\frac{2\Gamma}{\Delta}\right)^2}{1+
\left(\frac{2\Gamma}{\Delta}\right)^2}\\
&\leq N^2\left(\frac{2\Gamma}{\Delta}\right)^2.
\end{split}
\end{equation}
Then, we do the integral and find
\begin{equation}
\begin{split}
p_X' - p_X &\leq \frac{N^2 4\Gamma^2}{\sigma\pi^{1/2}}\int_{-\infty}^{\infty}d\Delta\,\frac{1-e^{-\Delta^2/\sigma^2}}{\Delta^2}\\
&= \frac{8\Gamma^2N^2}{\sigma^2} = O\left(\left(\frac{N\Gamma}{\sigma}\right)^2\right).
\end{split}
\end{equation}
\bibliography{reference}
\end{document}